\documentclass{aa}

\usepackage{colordvi}
\def\r[#1]{\Red{#1}}

\usepackage{txfonts}

\usepackage{graphicx, epsfig}

\usepackage{natbib}
\bibpunct{(}{)}{;}{a}{}{,} 

\usepackage{multirow}

\def\5pt{\hspace{5pt}}

\usepackage{dcolumn}
\newcolumntype{.}{D{.}{.}{2}}
\newcolumntype{d}[1]{D{,}{\,\pm\,}{#1}}
\newcolumntype{t}[1]{D{*}{\rm\ to\ }{#1}}

\usepackage{url}

\begin{document}


\title{Space-VLBI polarimetry of the BL\,Lac object S5 0716+714: Rapid
polarization variability in the VLBI core}

\author{U. Bach\inst{1,2}
	\and T.P. Krichbaum\inst{1}
	\and A. Kraus\inst{1}
	\and A. Witzel\inst{1}
	\and J.A. Zensus\inst{1}}
\authorrunning{U. Bach et al.} 
\titlerunning{Space-VLBI polarimetry of the BL\,Lac object S5\,0716+714} 
\institute{Max-Planck-Institut f\"ur Radioastronomie, Auf dem H\"ugel 69, 53121 Bonn, Germany
\and
INAF - Osservatorio Astronomico di Torino, Via Osservatorio 20, 10025 Pino
Torinese, Italy}
\offprints{U. Bach, \email{bach@to.astro.it}}
\date{Received 29 July 2005; accepted 11 November 2005}

\abstract{

To determine the location of the intra-day variable (IDV) emission region within
the jet of the BL\,Lac object S5\,0716+714, a multi-epoch VSOP polarization
experiment was performed in Autumn 2000. To detect, image, and monitor the short
term variability of the source, three space-VLBI experiments were  performed
with VSOP at 5\,GHz, separated in time by six days and by one day.
Quasi-contemporaneous flux density measurements with the Effelsberg 100\,m radio
telescope during the VSOP observations revealed variability of about 5\,\% in
total intensity and up to 40\,\% in linear polarization in less than one day.
Analysis of the VLBI data shows that the variations are located inside the VLBI
core component of 0716+714. In good agreement with the single-dish measurements,
the VLBI ground array images and the VSOP images, both show a decrease in the
total flux density of $\sim 20$\,mJy and a drop of $\sim 5$\,mJy in the linear
polarization of the VLBI core. During the observing interval, the polarization
angle rotated by about 15 degrees. No variability was found in the jet. The high
angular-resolution VSOP images are not able to resolve the variable component
and set an upper limit of $<0.1$\,mas to the size of the core component. From
the variability timescales we estimate a source size of a few micro-arcseconds
and brightness temperatures exceeding $10^{15}$\,K. We discuss the results in
the framework of source-extrinsic (interstellar scintillation induced) and
source-intrinsic IDV models.

}
%
\maketitle


\section{Introduction}

Since the discovery of intra-day variability (IDV, i.e. flux density and
polarization variations on time scales of less than 2 days) about 20 years ago
(\citealt{1986MitAG..65..239W,1987AJ.....94.1493H}), it has been shown that IDV
is a common phenomenon among extra-galactic compact flat-spectrum radio sources.
It is detected in a large fraction of this class of objects (e.g.,
\citealt{1992A&A...258..279Q,2001MNRAS.325.1411K,2003AJ....126.1699L}). The
occurrence of IDV appears to be correlated with the compactness of the VLBI
source structure on milliarcsecond scales: IDV is more common and more
pronounced in objects dominated by a compact VLBI core than in sources that show
a prominent VLBI jet (\citealt{1992A&A...258..279Q}; see also
\citealt{2001ApJ...554..964L}). In parallel to the variability of the total flux
density, variations in the linearly polarized flux density and the polarization
angle have been observed in many sources (e.g.,
\citealt{1989A&A...226L...1Q,1999bllp.conf...49K,1999NewAR..43..685K,2003A&A...401..161K,2004ChJAA...4...37Q}).
Both correlations (e.g.\ in 0716+714, \citealt{1996AJ....111.2187W}) and
anti-correlations (e.g.\ in 0917+624, \citealt{2002ChJAA...2..325Q}) between the
total and the polarized flux density are observed. In many cases the IDV
phenomenon is explained and modelled using refractive interstellar scintillation
(RISS, e.g., \citealt{1995A&A...293..479R,2001Ap&SS.278....5R}).  On the other
hand some effects remain that cannot be easily explained by interstellar
scintillation and that probably demand an explanation via source intrinsic 
relativistic jet physics. (e.g.\
\citealt{1996ChA&A..20...15Q,2002ChJAA...2..325Q,2004ChJAA...4...37Q}). There
are also sources like 0716+714 and 0954+658, where correlated intra-day
variability between radio and optical wavelengths is observed, which suggests
that at least part of the observed IDV has a source-intrinsic origin (e.g.,
\citealt{1990A&A...235L...1W,1991ApJ...372L..71Q,1996AJ....111.2187W}). We also
note that the recent detection of IDV at millimetre wavelengths in 0716+714
(\citealt{2002PASA...19...14K,2003A&A...401..161K,Agudo2006}) also poses a
problem for the interpretation of IDV by interstellar scintillation.

Independent of the physical cause of IDV (source intrinsic, or induced by
propagation effects), it is obvious that IDV sources must contain one or more
ultra-compact emission regions. Using scintillation models, typical source sizes
of a few ten micro-arcseconds have been derived (e.g.,
\citealt{1995A&A...293..479R,2002Natur.415...57D,2003ApJ...585..653B}). In the
case of source intrinsic variability and when using the light-travel-time
argument, even smaller source sizes of a few micro-arcseconds are obtained. In
this case it implies apparent brightness temperatures of up to $10^{18-19}$\,K
(in exceptional cases up to $10^{21}$\,K), far in  excess of the inverse Compton
limit of $10^{12}$\,K (\citealt{1969ApJ...155L..71K}). These high apparent
brightness temperatures can be reduced by relativistic beaming with  high
Doppler-factors (e.g.,
\citealt{1991A&A...241...15Q,1996ChA&A..20...15Q,2002PASA...19...77K}). At
present it is unclear if Doppler-factors larger than 50 to 100 are even possible
in compact extragalactic radio sources. 

One of the motivations of this VLBI monitoring, therefore, was to find out where
the IDV component is located in the jet and to directly search for related rapid
structural variability on milliarcsecond- to sub-milliarcsecond-scales. 
Previous observations with ground-based VLBI has already suggested the presence
of polarization IDV in a number of objects. In the case of 0917+624 and
0954+658, the IDV could be related to a component inside, or very close to the
VLBI core (\citealt{2000MNRAS.315..229G}). For 0716+714, however, it has been
claimed that the variable component is located at about $\sim25$\,mas separation
from the VLBI core  (\citealt{2000MNRAS.313..627G}). 

The BL\,Lac object S5 0716+714 is one of the brightest BL\,Lac objects  in the
sky. Only a lower limit to its redshift is known ($z >0.3$,
\citealt{1996AJ....111.2187W}, and references therein). The source 0716+714 is
also one of the best studied IDV sources, showing rapid variability in every
wavelength band where it was observed (e.g.\ \citealt{1996AJ....111.2187W}). In
the radio its linear polarization can vary up to a factor of two, often
accompanied by rotation of the polarization angle
(\citealt{1995ARA&A..33..163W}  and ref. therein). A direct conclusion drawn
from such polarization angle variations is the presence of multiple
sub-components, each exhibiting different compactness and polarization that
interact either due to deflection in the interstellar medium or to
superposition of multiple components inside the beam
(\citealt{1993ChA&A..17..229Q,1995ChA&A..19...69Q,2001Ap&SS.278..129R}).

Intraday variability is most pronounced at cm-wavelengths where the resolution
of ground based radio-interferometers is limited to the mas-scale. A factor of 3
to 4 higher angular resolution now is provided with the VLBI Space Observatory
Program (VSOP, \citealt{1998Sci...281.1825H,2000PASJ...52..955H}). This offers
a much closer look at the origin of the rapid variability in 0716+714. To
search for structural variability on time-scales of hours to a few days, we
observed 0716+714 in a multi-epoch VSOP polarization VLBI experiment in
September and October 2000. In this paper we will present and discuss the
detected rapid total intensity and polarization variations and compare them with
similar variations seen on VLBI scales with space-VLBI.

Throughout this paper we use a flat universe with the
following parameters: a Hubble constant of $H_0=71$\,km\,s$^{-1}$\,Mpc$^{-1}$,
a pressure-less matter content of $\Omega_{\rm m}=0.3$, and a cosmological
constant of $\Omega_{\rm \lambda}=0.7$ (\citealt{2003ApJS..148..175S}).

The observations and the data reduction procedures will be described in
Sect.~\ref{VSOP_sec:observations}. This section will be followed by the
presentation of the results (Sect.~\ref{VSOP_sec:results}) and the discussion of
our findings (Sect.~\ref{VSOP_sec:discussion}). At the end we give a
summary with the conclusions (Sect.~\ref{VSOP_sec:conclusions}).

\section{Observations and data reduction}\label{VSOP_sec:observations}

\subsection{VLBI data}
\begin{figure*}[t]
\centering
\includegraphics[angle=0,width=17cm] {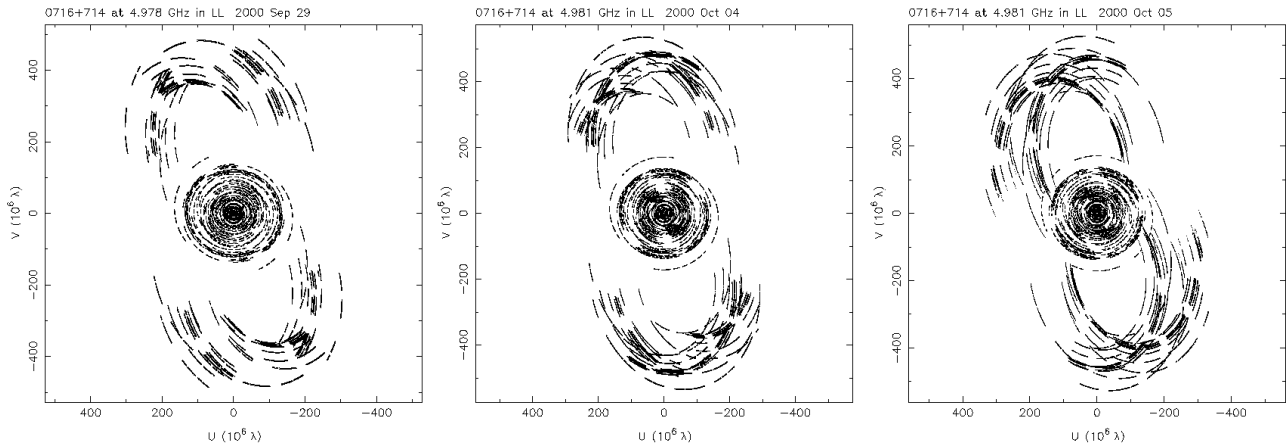}
   \caption{$(u,v)$ coverage of 0716+714 at
   September 29 (left), October 4 (middle), October 5 (right). The individual
   coverage are nearly identical, which allows a reliable comparison of the
   obtained maps.}
   \label{VSOP_fig:uvplots} 
\end{figure*}

An array of 12 antennas consisting of the 10 stations of NRAO's VLBA, the 100\,m
radio telescope of the Max-Planck-Institut f\"ur Radioastronomie in Effelsberg
(Germany), and the 8\,m HALCA antenna of the VSOP was used to follow the
short-term variability of 0716+714 at 5\,GHz within one week. The source was
observed at 5 GHz during three epochs, namely September 29 (project code:
V053\,A2), October 4 (V053\,A4), and October 5 2000 (V053\,A5). We note that a
total intensity VSOP image of 0716+714 from these experiments  was used already
in another publication, where we study the long-term jet kinematics of the
sources (\citealt{2005A&A...433..815B}, hereafter B05). Each of the three 16\,h
VSOP observations resulted in dense and nearly identical ($u,v$)-coverages,
which are shown in Fig.~\ref{VSOP_fig:uvplots}. The details of the observations
are summarised in Table~\ref{VSOP_tab:obslog}. The sources 0836+710 and 0615+820
were used as calibrators for the ground array to check the stability of the
amplitude calibration between the epochs and to calibrate the polarization data.
The calibrators were not observed by the HALCA antenna, because it is not
possible to steer HALCA to different source positions during one observation.

\begin{table}[htbp]
\caption{Observing log. The total
integration time, mutual observing time with HALCA, total flux density, the
uniform weighted beam size, and the residual noise are given for each epoch and
each source.}
\label{VSOP_tab:obslog}
\begin{minipage}{\linewidth}
\renewcommand{\footnoterule}{}
\renewcommand{\thempfootnote}{\textit\alph{mpfootnote}}
\centering
\scriptsize
\begin{tabular}{llccccc}
\hline
Epoch &
Source &
\multicolumn{1}{c}{Int.} &
\multicolumn{1}{c}{HALCA} &
\multicolumn{1}{c}{$S_{\rm tot}$} &
\multicolumn{1}{c}{Beam, P.A.} &
\multicolumn{1}{c}{$S_{\rm rms}$}\\
 &
 &
\multicolumn{1}{c}{[h]} &
\multicolumn{1}{c}{[h]} &
\multicolumn{1}{c}{[Jy]} &
\multicolumn{1}{c}{[${\rm mas\times mas}$], [$^\circ$]} &
\multicolumn{1}{c}{[mJy]}\\
\hline
\multirow{4}{22pt}{Sep.~29\footnote{In this epoch the Pie Town antenna was
substituted by a single VLA antenna.}} 
                       & \multirow{2}{11pt}{0716+714\footnote{Two lines are given for 0716+714:
the first gives the parameters including HALCA and the second for the ground array data alone.}} & 7.67 & 2.75 & 0.55 & $0.58\times0.27$, $-80$& 0.492\\ 
                       &          & 7.67 & 0.00 & 0.58 & $0.95\times0.78$, $-26$& 0.034\\
                       & 0836+710 & 1.17 & 0.00 & 2.23 & $1.10\times0.88$, $-10$& 0.193\\
                       & 0615+820 & 1.17 & 0.00 & 0.74 & $0.90\times0.78$, $-42$&\ 0.191
\vspace{1mm}\\
\multirow{4}{22pt}{Oct.~4} & \multirow{2}{11pt}{0716+714$^b$} & 7.72 & 4.59 & 0.52 & $0.58\times0.23$, $-79$& 0.513\\
                       &          & 7.72 & 0.00 & 0.55 & $0.92\times0.77$, $-25$& 0.041\\
                       & 0836+710 & 1.20 & 0.00 & 2.28 & $0.99\times0.83$, $-46$& 0.168\\
                       & 0615+820 & 1.19 & 0.00 & 0.75 & $0.84\times0.82$, $-47$&\ 0.173
\vspace{1mm}\\
\multirow{4}{22pt}{Oct.~5} & \multirow{2}{11pt}{0716+714$^b$} & 7.92 & 5.47 & 0.52 & $0.58\times0.24$, $-71$& 0.580\\
                       &          & 7.92 & 0.00 & 0.56 & $0.85\times0.81$, $-28$& 0.038\\
                       & 0836+710 & 1.21 & 0.00 & 2.27 & $0.96\times0.81$, $-74$& 0.175\\
                       & 0615+820 & 1.23 & 0.00 & 0.76 & $0.93\times0.76$, $-55$& 0.171\\
\hline
\end{tabular}
\end{minipage}
\end{table}

The 100\,m RT at Effelsberg not only participated as a VLBI antenna, but also
was used to monitor the flux density and polarization variability of all
programme sources. These flux density measurements were made in gaps between
VLBI scans using pointing cross-scans in azimuth and elevation. From this and
the flux and polarization measurements of primary calibrators (3C\,286, 3C\,48
and 3C\,295) before and after the VLBI experiments, the orientation of the
polarization E-vector (EVPA, Electric Vector Position Angle) was also obtained (see
Sect.~\ref{sec:effred} for details).

The data were recorded in VLBA format with two 16\,MHz baseband channels per
circular polarization in two-bit sampling, resulting in a recording rate of
128\,Mbps at the ground array stations. The HALCA antenna observed only LCP and
the data was recorded at several tracking stations (typically 3 per epoch). The
correlation was done at the VLBA correlator in Socorro, NM.

Although HALCA is only able to observe in left circular polarization (LCP), 
VLBI polarimetry becomes possible, if enough other stations of the remaining
VLBI array measure both circular polarizations. It is possible to
cross-correlate LCP from HALCA with RCP from the ground array stations and thus
obtain the cross-polarized correlations on the space baselines. Normally one
needs both cross polarizations, $RL$ and $LR$, to image the linear polarization
of a source; but using a complex $Q+iU$ image, it is possible to include
antennas with only a single cross polarization (e.g., AIPS Memo 79, 1992,
W.D.~Cotton)\footnote{\url{http://www.aoc.nrao.edu/aips/aipsmemo.html}}. Further
details about polarization observations using HALCA are given in, e.g.,
\cite{1999NewAR..43..691G}, \cite{2000PASJ...52.1055K}, and
\cite{2001MNRAS.320L..49G}.

The post-correlation analysis was done using NRAO's Astronomical Image
Processing System ({\sc Aips}). After loading the data into {\sc Aips}, the
standard amplitude and phase calibrations were performed. Since HALCA does not
provide pulse calibration information, a manual phase calibration was done to
remove offsets between the two intermediate frequency channels (IFs).

\begin{figure}[htbp]
\centering
\includegraphics[angle=0,width=8cm]
{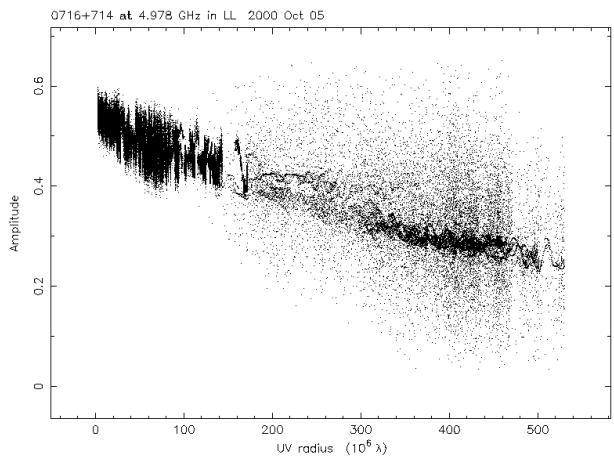}
\caption{Amplitude vs.
$(u,v)$-distance of 0716+714 (Oct. 5). Clearly visible is the higher noise at
the HALCA baselines at large $(u,v)$-distances. The dotted line (marginally
visible in the centre of the HALCA data) represents the final 
clean model.}\label{VSOP_fig:radplot}  
\end{figure}

At this point the data were exported as $(u,v)$-FITS files (task FITTP) from
{\sc Aips} to {\sc Difmap} (\citealt{1994BAAS...26..987S}), where the imaging,
phase, and amplitude calibration was done, using the CLEAN
(\citealt{1974A&AS...15..417H}) and SELFCAL procedures. Because of the small
diameter of the HALCA antenna, the data from the ground array stations yield a
much better signal-to-noise ratio (Fig.~\ref{VSOP_fig:radplot}). Therefore, we
first imaged the ground array data alone to obtain a good initial Stokes $I$
image. After this, we included the HALCA antenna in the imaging and
self-calibration  process, using a strong Gaussian taper (10\,\% at
450\,M\,$\lambda$) during the first iterations and subsequently decreasing the
taper. All images were obtained using uniform weighting. For the ground array
images the gridding weights were scaled by the amplitude errors raised to the
power $-2$, which favours high quality data with small errors, and for the VSOP
images equal weights were applied to achieve highest possible angular
resolution. The amplitude errors were derived by the scattering of the data when
we averaged the data sampled at 4-second intervals to 60-second intervals. 

The self-calibration was done in total intensity in steps of several
phase-calibrations followed by careful amplitude calibration. During the
iteration process, the solution interval of the amplitude self-calibration was
shortened from intervals as long as the whole observational time (in the initial
imaging steps) down to minutes at the end of the imaging process. Finally the
self-calibrated data were  reimported into {\sc Aips} (using task FITLD),
where the calibration of the leakage terms (D-terms) and the polarization
imaging was done.

\subsubsection{Feed calibration}\label{VSOP_sec:dtermcal}

Calibration of the instrumental polarization and the leakage terms (D-terms)
from the left to the right circular polarization feeds at each antenna was done
using the task LPCAL in {\sc Aips} (for the method see:
\citealt{1995AJ....110.2479L}). Therefore the total-intensity model produced
with IMAGR was split into sub-models using the task CCEDT. The model was
separated into three to four sub-models that should represent a small number of
``isolated components'', with clearly defined individual polarization
properties, suitable for determining the instrumental cross-polarization with
LPCAL. This was done in several steps and with different sub-models, to ensure
that the D-terms do not depend critically on the choice of the different source
and  sub-models.

\begin{figure}[htbp]
\centering
\includegraphics[angle=0,width=9cm] {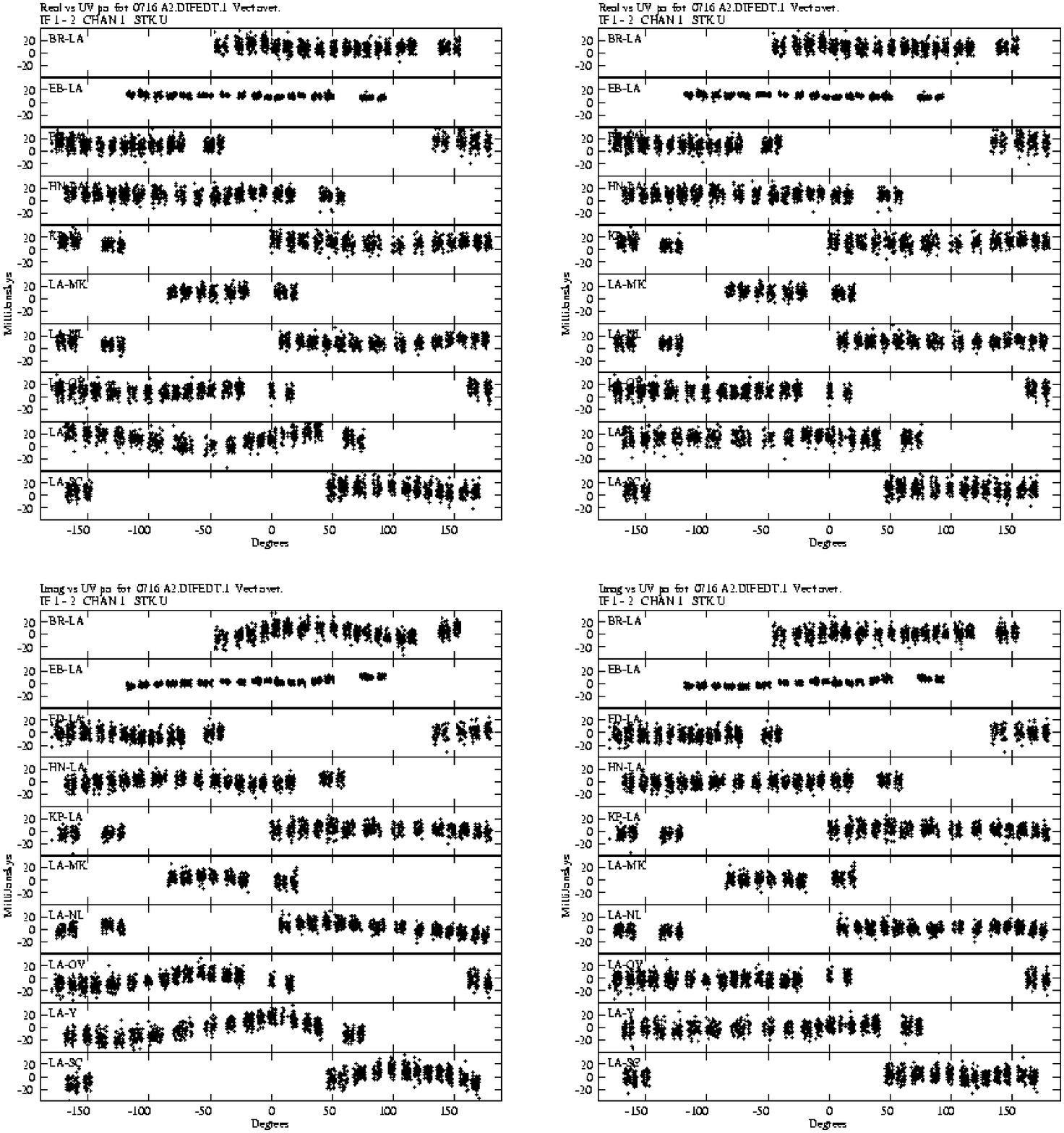}
   \caption[Stokes $U$ real and imaginary data vs. P.A.]{An example for the
   tests made to check the reliability of the D-term solutions: the panels show
the real (top) and imaginary part (bottom) of the Stokes $U$ data on all ground
array baselines to Los Alamos plotted versus the parallactic angle (Sep. 29).
{\bf Left panels:} Before application of D-term calibration. {\bf Right panels:}
After the D-term calibration was applied. The calibration successfully removes
the systematic variation of the amplitude with parallactic
angle.}\label{VSOP_fig:real_imag-pa}
\end{figure}

The LPCAL algorithm is robust against different source structures, nevertheless
one should prefer core-dominated sources for the calibration. All sources in
our observations meet this criterion. The VLBA antennas and Effelsberg exhibit
D-term values between 0.5\,\% and 2\,\%. For HALCA we derived about $4$\,\%,
similar to what was found by previous observations (e.g.,
\citealt{1999NewAR..43..691G}). A number of feed solutions were calculated by
using slightly different source models and/or different calibrator sources
(0816+710 and 0615+820). The RMS differences between these D-terms are typical
smaller than 0.3\,\% for the ground array stations and reach $\sim0.5$\,\% for
the HALCA satellite. The D-terms of HALCA could only be checked with different
source models of 0716+714, since the calibrators were not observed by the
satellite. Plots of the real versus imaginary cross-hand polarization data
indicated that a satisfactory D-term solution was obtained. This was also
verified in plots of the real and imaginary cross-hand data versus $(u,v)$
parallactic angle. After applying the D-term solution, the variations in the
visibility amplitudes with parallactic angle were removed. In
Fig.~\ref{VSOP_fig:real_imag-pa} we illustrate this, showing some
visibilities before and after application of the D-term calibration.

To estimate how the D-term variations affect the source polarization images,
fractional polarization maps with different feed calibration tables were made
and then subtracted. The residual amount of fractional polarization in the
subtracted images can be used to measure the uncertainty introduced by the
different calibrations. This yields an accuracy of $\Delta P/P = 0.1$\,\% in
fractional polarization for the VLBI core of 0716+714 and $\Delta P/P = 1.2$\,\%
for the weaker jet. The effect of different D-terms on the accuracy of the EVPA
was measured in a similar way. Here we obtained an uncertainty of the
orientation on the E-vector of $\Delta {\rm EVPA}= 0.3^\circ$ for the core and
$\Delta {\rm EVPA}= 2.0^\circ$ for the jet.

\subsubsection{Polarization imaging}\label{VSOP_sec:polimag}

The usual imaging technique for linear polarization is to form  $Q$ and $U$
values from the observed $RL$ and $LR$ correlations and separately image and
deconvolve the $Q$ and $U$ images. The {\sc Aips} task COMB can be used to
combine the $Q$ and $U$ images to obtain a linear polarization intensity map
($I_{\rm P}=\sqrt{Q^2+U^2}$) and a map that represents the electric vector
polarization angle ($EVPA=\chi=0.5\,\arctan{\frac{U}{Q}}$). But this is only
possible in observations where both cross-polarized correlations, $RL$ and
$LR$, are available for each interferometer baseline.

Since the HALCA antenna observed only in LCP, only one of the two cross
polarized correlations is available for the space-baselines.  Nevertheless, it
is still possible to form a linear polarization image: in this case complex
imaging with a complex deconvolution of $Q+iU$ has to be applied. The {\sc Aips}
software package also offers these tasks. We used the procedure CXPOLN to build
the complex image and beam, along with the task CXCLN, which does the complex
cleaning and which provides the cleaned $Q$ and $U$ images. These images were
then combined with COMB in the usual way to obtain linear polarization intensity
and polarization  angle images.

\subsubsection{EVPA calibration}\label{VSOP_sec:EVPA}

For the EVPA calibration we used Effelsberg measurements of the quasar 0836+710,
orienting its E-vector to ${\rm P.A.}= 106.3\,^\circ$, previously determined
from measurements relative to the primary calibrator 3C\,286. We further assumed
that between the angular scales covered by the Effelsberg beam and the VLBI
scale, there is no dominant polarized component that could affect the EVPA
calibration of the VLBI data. The existing VLA maps of 0836+710 support this
assumption, as does the comparison of the total intensity and linear
polarization measurements at Effelsberg
(Tables~\ref{VSOP_tab:ebsources}~\&~\ref{VSOP_tab:ebpol}) with the flux density
seen with the VLBA (Table~\ref{VSOP_tab:obslog}~\&~\ref{VSOP_tab:0836pol}).

\begin{table}[hbtp]
\centering
\caption{Comparison of the polarization properties of 0836+710 on single-dish ($P_{\rm
Eff}$ and $\chi_{\rm Eff}$) and VLBI ($P_{\rm VLBI}$ and $\chi_{\rm VLBI}$)
scales. Using the fixed EVPA from Effelsberg, the measured EVPA from VLBI was
corrected by $\Delta \chi = \chi_{\rm Eff} -  \chi_{\rm VLBI}$. The different
value for the correction in the last epoch is due to the choice of a different
reference antenna during the calibration.} 
\label{VSOP_tab:0836pol}
\scriptsize
\begin{tabular}{ld{2}d{3}d{3}d{3}d{4}}
\hline
Epoch & 
\multicolumn{1}{c}{$P_{\rm Eff}$} &
\multicolumn{1}{c}{$\chi_{\rm Eff}$} &
\multicolumn{1}{c}{$P_{\rm VLBI}$} &
\multicolumn{1}{c}{$\chi_{\rm VLBI}$} &
\multicolumn{1}{c}{$\Delta \chi$} \\
&
\multicolumn{1}{c}{[mJy]} &
\multicolumn{1}{c}{$[^\circ]$} &
\multicolumn{1}{c}{[mJy]} &
\multicolumn{1}{c}{$[^\circ]$} &
\multicolumn{1}{c}{$[^\circ]$} \\
\hline
29 Sep & 160, 1 & 106.3 , 0.3 & 131 , 13 & 125.8,1.2 &-19.5 , 1.2\\
04 Oct & 160, 1 & 106.3 , 0.3 & 133 , 13 & 125.6,1.0 &-19.3 , 1.0\\
05 Oct & 160, 1 & 106.3 , 0.3 & 128 , 13 & 96.0,1.1  & 10.3 , 1.1\\
\hline
\end{tabular}
\end{table}

\subsection{Effelsberg measurements}\label{sec:effred}

In gaps between VLBI scans, we measured the flux density and polarization of
the program sources and calibrators. The measurements were performed using
standard pointing cross-scans with slews in azimuth and elevation. In addition
to the VLBI targets, we observed 0951+699 and 3C\,286 as additional flux density
and polarization calibrators. In Table~\ref{VSOP_tab:ebsources} we summarise
the mean flux density (col. 2) for each source, the number of pointing scans
(col. 3), the modulation index $m=100 \cdot S/\langle S\rangle $, with the mean flux density
$\langle S\rangle $ (col. 4), and the reduced chi-square testing the assumption of non-variability. 

\begin{table}[htbp]
\caption{Effelsberg observing log. Shown are the mean
flux density, $\langle S\rangle $, the number of scans, $N$, the modulation index, $m$, and
the reduced $\chi^2_{\rm r}$ from a linear fit (see text for details).}
\centering
\label{VSOP_tab:ebsources}
\begin{tabular}{ld{8}rcc}
\hline
\multicolumn{1}{c}{Source} &
\multicolumn{1}{c}{$\langle S\rangle $ [Jy]} &
\multicolumn{1}{c}{$N$} &
\multicolumn{1}{c}{$m$ [\%]} &
\multicolumn{1}{c}{$\chi^2_{\rm r}$}\\
\hline
0615+820\,$^a$ &  0.773 , 0.003 & 35 & 0.42 & 0.08\\
0716+714\,$^a$ &  0.747 , 0.018 & 31 & 2.38 & 2.25\\
0836+710\,$^a$ &  2.500 , 0.008 & 45 & 0.33 & 0.05\\
0951+699       &  3.367 , 0.005 &  4 & 0.14 & 0.01\\
3C\,286        &  7.521 , 0.074 & 11 & 0.98 & 0.42\\
3C\,295        &  6.593 , 0.056 &  8 & 0.85 & 0.31\\
3C\,48         &  5.537 , 0.019 &  2 & 0.34 & 0.05\\
\hline
\multicolumn{5}{l}{\footnotesize $a$: target for VLBI} \\
\end{tabular}
\end{table}

A detailed description of the Effelsberg data reduction procedures is given in
\cite{2003A&A...401..161K}. In this experiment we used 3C\,286, 3C\,295, and
3C\,48 as the main flux density calibrators ($S_{\rm 3C\,286}=7.50$\,Jy, $S_{\rm
3C\,295}=6.59$\,Jy and $S_{\rm 3C\,48}=5.63$\,Jy) and 0836+710 as a secondary
calibrator, which was observed adjacent to each scan on 0716+714. Also 0615+820
was observed regularly as a secondary calibrator and is known to be non-variable
at least on time-scales of days (\citealt{Kraus1997}). The linear polarization
measurements were calibrated using 0836+710 and 3C\,286, which both have known
polarization properties, as well as 0951+699 which is unpolarized, and all only
vary over longer time-scales. After calibration, the residual error of the
Effelsberg EVPA was $\sim0.3^\circ$.

\section{Results}\label{VSOP_sec:results}

In this section the total intensity and linear polarization maps of 0716+714
from the ground array data and the VSOP data are presented and analysed. The
high dynamic range of the ground array data (peak/RMS $\sim 15\,000$) enables us
to follow and study the weak jet up to $\sim 12$\,mas core separation. The three
times higher resolution in the VSOP images allows a detailed study of the core
structure and its variability. The results and analysis of the total intensity
and polarization measurements with the 100\,m RT in Effelsberg are presented in
Sect.~\ref{VSOP_sec:Eff}. We would like to note that during each of the three
VSOP observations, 0716+714 showed only moderate variations in total intensity
($\sim 5$\,\%, peak to peak during one epoch). A more pronounced variable
source would violate the principle of stationarity in aperture synthesis and
would lead to a severe degradation of the reconstructed interferometric image.
Here, these effects are small and can be neglected.

Imaging simulations performed by \cite{Hummel1987} have shown that intra-day
variability during VLBI observations mainly reduces the achievable dynamic range
due to residual side lobes in the images without changing the source structure.
In these simulations even larger variations ($\sim 80$\,\%) were considered.
Intensity variations of about 5\,\%, which are present in our observations, are
comparable to the usual uncertainties of the $T_{\rm sys}$ measurements ($\sim
5-10$\,\%), which are used to calibrate the visibility amplitudes. The final
images after self-calibration show a nearly constant noise level outside the
source structure without any symmetric features indicating that there are no
residual side lobes present. Therefore, we can neglect the effects of the IDV on
our VLBI maps. The larger variability of the linear polarization (up to $\sim
40$\,\%) probably degrades the linear polarization images to some extent, but
should not severely affect the self-calibration procedure, because the phase and
amplitude self-calibration of  $LL$, $RR$, $LR$, and $RL$ during the imaging
process was done using the total intensity data. 

\subsection{Ground array data}\label{VSOP_sec:groundarray}
\begin{figure*}[htbp]
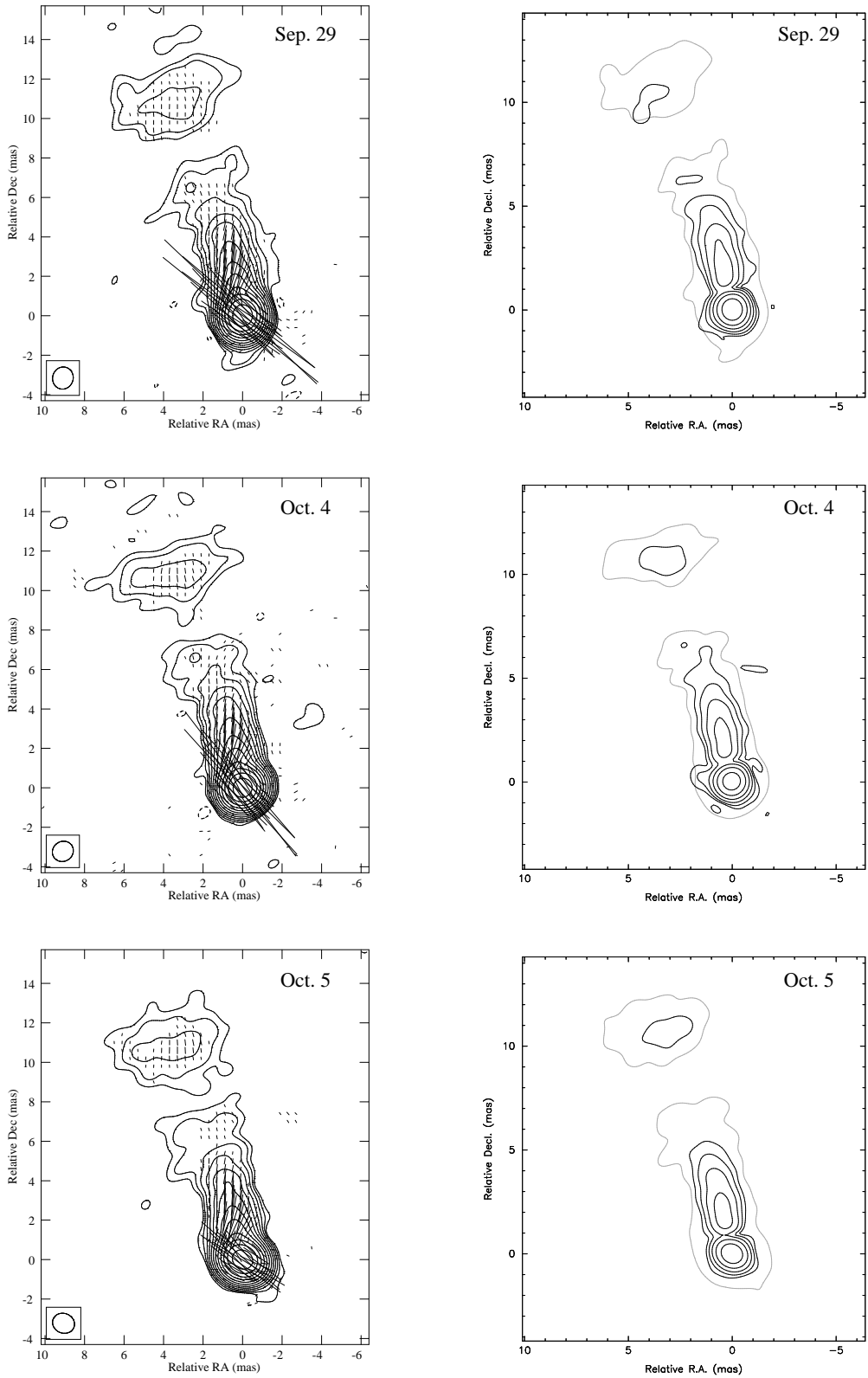

 \hbox{\hspace*{15mm}
 \includegraphics[angle=0,width=5.9cm] {3943_fig7.ps}\hspace*{15mm}
 \includegraphics[bb=41 140 560 775,angle=0,width=5.9cm,clip] {3943_fig8.ps}
 \put(-245,175){\makebox(0,0){Sep. 29}}
 \put(-25,175){\makebox(0,0){Sep. 29}}}
 \hbox{\hspace*{15mm}
 \includegraphics[angle=0,width=5.9cm] {3943_fig9.ps}\hspace*{15mm}
 \includegraphics[bb=41 140 560 775,angle=0,width=5.9cm,clip] {3943_fig10.ps}
 \put(-245,175){\makebox(0,0){Oct. 4}}
 \put(-25,175){\makebox(0,0){Oct. 4}}}
 \hbox{\hspace*{15mm}
 \includegraphics[angle=0,width=5.9cm] {3943_fig11.ps}\hspace*{15mm}
 \includegraphics[bb=41 140 560 775,angle=0,width=5.9cm,clip] {3943_fig12.ps}
 \put(-245,175){\makebox(0,0){Oct. 5}}
 \put(-25,175){\makebox(0,0){Oct. 5}}}
   \caption{{\bf Left panels:} Ground array contour
   maps of Stokes $I$ of 0716+714 with polarization vectors superimposed. The
   length of the polarization vectors is proportional to the intensity of the
   linear polarization (1\,mas corresponds to 2.5\,mJy/beam in $P$). Contours
   start at 0.13\,mJy/beam, increasing in steps of 2. {\bf Right panels:}
   Contour maps of the linear polarization. Contours start at 0.25\,mJy/beam,
   increasing in steps of 2. The lowest contour of the total intensity is given
   by the outer grey line. Time order is from top to bottom: 29 Sep, 4 Oct, and 5
   Oct 2000. The beam size, total flux density, and RMS are given in the observing
   log, Table~\ref{VSOP_tab:obslog}.}\label{VSOP_fig:groundmaps}
\end{figure*}

\begin{figure}[htbp]
\centering
\includegraphics[bb=100 40 490 830,angle=-90,width=9cm,clip]
{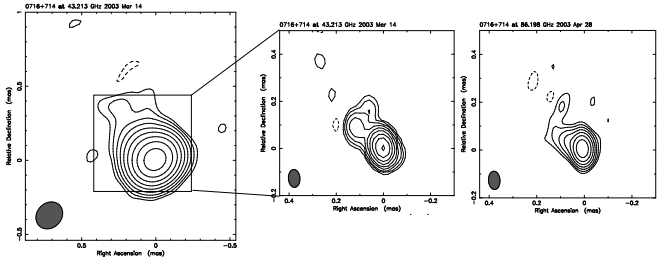}
\caption{Contour images of 0716+714
at 43\,GHz and 86\,GHz showing that the jet is bent by about $20^\circ$ in the
inner 1\,mas with respect to the jet at lower resolution. {\bf Left:}
43\,GHz VLBA contour map at a resolution of $166 \times  196\,\mu$as at 
$-43.5^\circ$ from March 14, 2003. The peak flux density is 2\,Jy/beam and the
contours start at 6\,mJy increasing in steps of 2. {\bf Middle:} Super
resolved ($76 \times  49\,\mu$as at $3.5^\circ$) version of the 43\,GHz map.
{\bf Right:} 86\,GHz map at a resolution of $76 \times  49\,\mu$as at
$3.5^\circ$ from April 28, 2003. Peak flux density is 2.4\,Jy/beam and the
contours start at 20\,mJy.}\label{VSOP_fig:mm_map} 
\end{figure}

\begin{figure}[htbp]
\centering
\includegraphics[bb=0 20 560 822,angle=0,width=9.3cm,clip] {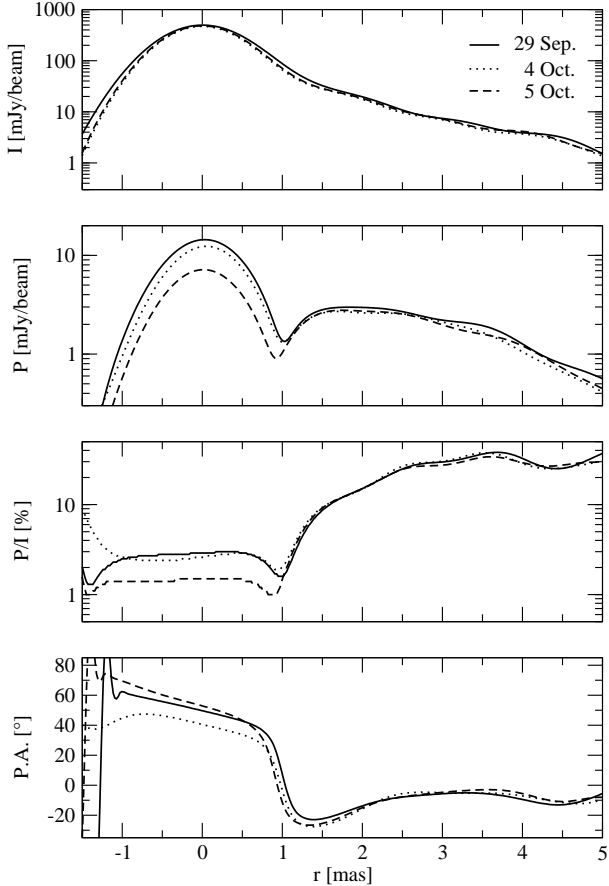}
   \caption{Slices along ${\rm P.A.}=12.5^\circ$ through the total intensity,
   linear polarization intensity, fractional polarization, and EVPA
   images of the core and the jet of 0716+714.}\label{VSOP_fig:groundprofile}
\end{figure}

The ground array images showing the parsec-scale structure of the jet of
0716+714 are shown in Fig.~\ref{VSOP_fig:groundmaps}. The panels in the left row
show the total intensity contours of 0716+714 with polarization E-vectors
superimposed for all three epochs. The right panels of the figure show
polarization  contour maps. The intensity maps reveal a one-sided core jet
structure with a north-oriented jet extending up to $\sim12$\,mas core
separation along P.A. $\approx15^\circ$. Between the core and the jet the EVPA
is misaligned by $\sim 60^\circ$. In the jet and on mas-scales, the electric
field vectors are aligned well with the jet axis. Provided that the jet emission
is optically thin, which is supported by the high degree of fractional
polarization (Fig.~\ref{VSOP_fig:groundprofile}), the alignment suggests that
the magnetic field is oriented perpendicular to the jet axis. The $60^\circ$
misalignment of the EVPA of the core can be either explained by Faraday
rotation, the fact that the core is optically thick, or by jet bending in the
inner most region near the core. High Faraday rotation has been found in many
AGN cores (e.g., \citealt{2004ApJ...612..749Z,2003ApJ...589..126Z} and
references therein) with the tendency towards lower rotation measures in BL\,Lac
objects than in quasars, but we are not aware of any rotation measures for the
core of 0716+714. However, recent mm-VLBI observations at 43\,GHz and 86\,GHz
show that on the 0.2\,mas scale, the jet is strongly curved, oriented along P.A.
$\approx50^\circ$ (Fig.~\ref{VSOP_fig:mm_map}). This supports the idea that the
EVPA orientation follows the bent jet down to the sub-mas scale and that even in
the vicinity of the core, the magnetic field is perpendicular to the jet axis.

Inspection of the three 6\,cm VLBI images in Fig.~\ref{VSOP_fig:groundmaps}
reveals no major changes in the total intensity structure between the epochs.
Table~\ref{VSOP_tab:obslog} shows a small and, at this stage, marginal decrease
of 5\,\% in the peak flux from the first to the last two epochs. A more detailed
parametrization of the maps, however, is given  by
Table~\ref{VSOP_tab:VLBAresult}. Here we show the integrated flux densities
separately for core and jet.

The measurements were made using the tasks IMSTAT and TVSTAT in {\sc Aips}, and
the errors are derived from the scatter between the individual measurements on
slightly different calibrated  maps and with different window sizes. The task
IMSTAT integrates over a rectangular region that is not necessarily in good
agreement with the source structure, whereas TVSTAT integrates over a polygonal
region that is specified by the user. We preferred to use IMSTAT and TVSTAT
instead of model-fitting Gaussian components to the images, since they also give
us more precise measurement of the extended emission.

Attempts to parameterise the total intensity and polarization images with model
fitting were focused on the problem that one either needs to use a different
number of components to fit $I$ and $P$ or to fix the component
position in the $I$ or $P$ images. In both cases one cannot be sure that all of
the extended jet emission is adequately represented by the model fit and, more
important, if measurements are comparable between the epochs.

\begin{table}[htbp]
\centering
\caption{Summary of the ground VLBI maps.}
\label{VSOP_tab:VLBAresult}
\begin{tabular}{lld{4}d{2}d{3}}
\hline
Epoch &  & \multicolumn{1}{c}{I [mJy]} & \multicolumn{1}{c}{P [mJy]} & 
\multicolumn{1}{c}{$\chi$ [$^\circ$]}\\
\hline
\multirow{2}{11pt}{29~Sep} & Core  & 520.3 , 26.9    & 12.1 , 1.3 &  49.4 , 4.1 \\
     & Jet   &  56.0 ,\5pt 4.7 &  7.4 , 0.8 & -10.8 , 5.6 \\
\multirow{2}{11pt}{04~Oct} & Core  & 499.3 , 26.1    & 11.8 , 1.3 &  40.7 , 4.0 \\
     & Jet   &  54.8 ,\5pt 6.3 &  7.3 , 0.8 & -11.2 , 7.8 \\
\multirow{2}{11pt}{05~Oct} & Core  & 503.9 , 25.4    &  6.5 , 1.1 &  52.7 , 5.2 \\
     & Jet   &  54.7 ,\5pt 6.0 &  7.5 , 0.8 &  -9.5 , 7.4 \\
\hline
\end{tabular}
\end{table}

From Table~\ref{VSOP_tab:VLBAresult} one can see that the core flux density
decreases from 520\,mJy to about 500\,mJy in the later epochs, whereas the jet
flux density stays relatively stable at $\sim 55$\,mJy. Using the technique of
amplitude self-calibration improves the quality of the VLBI images by applying
antenna-based gain corrections to remove residual calibration errors that were
not corrected by the a priory amplitude calibration. Provided that the
signal-to-noise ratio is sufficiently high, after several iterations the data
and the model should converge around an average value set by the a priory
amplitude calibration. In these experiment time-dependent station gain
corrections of a few percent (typically $\sim 5$\,\%) were applied. Since the
procedure used in all epochs was the same, the relative accuracy of the flux
densities after self-calibration should be better than the absolute calibration
and can be measured by using the flux of the extended jet emission seen in each
experiment. Due to its parsec-scale size, the flux of the integrated jet
emission should not be variable on time-scales of days. From
Table~\ref{VSOP_tab:VLBAresult} we obtain 56, 54.8, and 54.7\,mJy for the
integrated jet flux for each epoch, respectively. Thus, we could attribute this
2.3\,\% variation between the individual measurements of the extended jet
emission to the remaining calibration uncertainty. While the integrated jet flux
remained constant at this level, the core flux varied by $\sim 20$\,mJy or $\sim
4$\,\%, a factor of two more than the jet.

In contrast to the marginal variability of the total flux density, strong
variability is seen in linear polarization. A visual inspection of the
polarization vectors in Fig.~\ref{VSOP_fig:groundmaps} indicates that the linear
polarization of the core component varies significantly. From the integrated
values in Table~\ref{VSOP_tab:VLBAresult}, one can see that there are no changes
(within the errors) in the polarized intensity between the first two epochs, but
in the last epoch the polarized intensity of the core drops by $\sim$45\,\%,
from $\sim 12.0$\,mJy to $\sim 6.5$\,mJy, corresponding to a decrease in
fractional polarization from $\sim2.4$\,\% to $\sim1.3$\,\%. As in total
intensity, the polarized intensity of the jet shows no variability and stays
very constant with a polarized flux of $7.4 \pm 0.1$\,mJy (or 1.3\,\% accuracy)
at an average degree of polarization of $\sim 13.6$\,\%. The average EVPA of the
jet is also constant at an angle of $-10.5 \pm 0.9^\circ$, whereas the core EVPA
rotates by $\sim 10-13^\circ$, from $\sim50^\circ$ to $\sim40^\circ$ and back to
$\sim53^\circ$ during the observations. 

Figure~\ref{VSOP_fig:groundprofile} summarises all these results. It shows
profiles of total intensity, linear polarization, degree of polarization, and
the orientation of the E-vector  plotted versus core separation. To avoid
confusion between the different profiles, we did not display the error bars in
the plots. The errors of the total intensity and the polarization are dominated
by the uncertainties of the amplitude calibration, about 2\,\%. The error of
the EVPA depends on the SNR of the linear polarization and varies between
$0.3\,^\circ$ for the VLBI core component and about $2\,^\circ$ in  the bright
parts of the jet (see also
Sects.~\ref{VSOP_sec:dtermcal}~\&~\ref{VSOP_sec:EVPA}). 

In summary, it seems that between the 3 observations no variation was seen in
the jet, neither in intensity nor in polarization. On the other hand, the total
flux density of the VLBI core component varied by about $4 \pm 2$\,\% in total
intensity and by about a factor of 2 in polarization. The polarization angle of
the jet remained constant within $\sim 2^\circ$, while the polarization angle of
the core varied by $\sim 10^\circ$. Therefore all variability appears in the
bright and unresolved VLBI core component of 0716+714.

Noticeable are the two peaks of up to 40\,\% fractional polarization in the jet
at $r\approx 2.5$\,mas, $r\approx3.5$\,mas, and $r\approx4.5$\,mas distance from
the core in Fig.~\ref{VSOP_fig:groundprofile}. At the same positions, B05 found
modelfit components in total intensity, which were used to study the kinematics in
the jet. These jet components apparently move superluminally, with
speeds\footnote{$H_0=71$\,km\,s$^{-1}$\,Mpc$^{-1}$, $\Omega_{\rm m}=0.3$ and
$\Omega_{\rm \lambda}=0.7$} of 6.9\,$c$ to 8.3\,$c$. This provides further
evidence that these are  real structures in the jet, probably shocks, where we
might have a higher density that enhances the total intensity and a more ordered
magnetic field that enhances the linear polarization. 

\subsection{Space VLBI data}\label{VSOP_sec:VSOP}

\begin{figure*}[htbp]
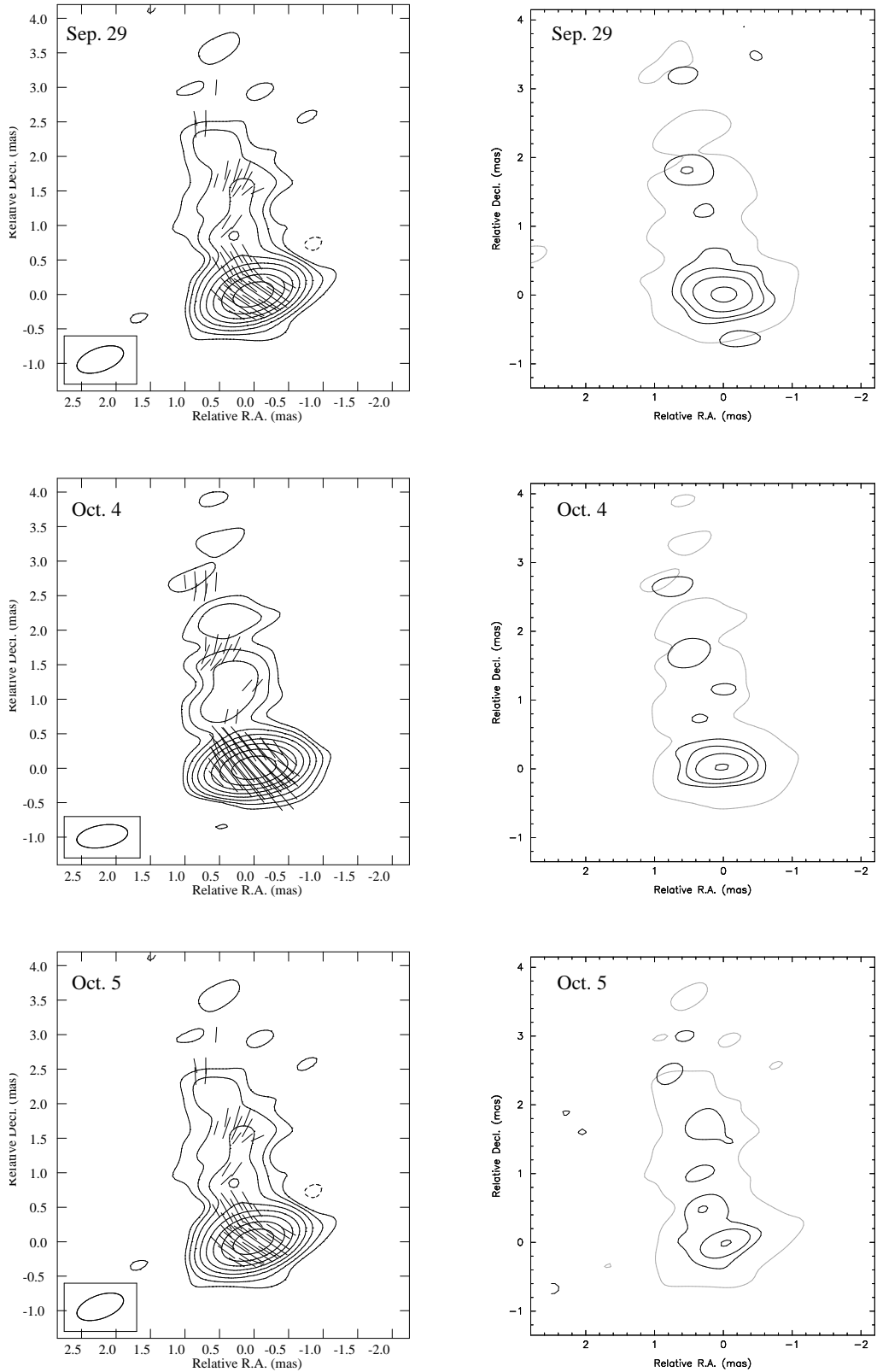

 \hbox{\hspace*{15mm}
 \includegraphics[bb=50 145 567 696,angle=0,width=6.4cm,clip] {3943_fig15.ps}\hspace*{10mm}
 \includegraphics[bb=41 140 560 775,angle=0,width=6cm,clip] {3943_fig16.ps}
 \put(-130,175){\makebox(0,0){Sep. 29}}
 \put(-345,175){\makebox(0,0){Sep. 29}}}
 \hbox{\hspace*{15mm}
 \includegraphics[bb=50 145 567 696,angle=0,width=6.4cm,clip] {3943_fig17.ps}\hspace*{10mm}
 \includegraphics[bb=41 140 560 775,angle=0,width=6cm,clip] {3943_fig18.ps}
 \put(-130,175){\makebox(0,0){Oct. 4}}
 \put(-345,175){\makebox(0,0){Oct. 4}}}
 \hbox{\hspace*{15mm}
 \includegraphics[bb=50 145 567 696,angle=0,width=6.4cm,clip] {3943_fig19.ps}\hspace*{10mm}
 \includegraphics[bb=41 140 560 775,angle=0,width=6cm,clip] {3943_fig20.ps}
 \put(-130,175){\makebox(0,0){Oct. 5}}
 \put(-345,175){\makebox(0,0){Oct. 5}}}
   \caption{{\bf Left panels:} VSOP contour maps of
   Stokes $I$ of 0716+714 with polarization vectors superimposed (1\,mas
   corresponds to 6.7\,mJy/beam). Contours start at 1.8\,mJy/beam and
   increasing in steps of 2. {\bf Right panels:} Contour maps of the linear
   polarization. Contours start at 1.3\,mJy/beam and increasing in steps of 2.
   The outline of the total intensity is indicated by the outer grey contour. Time
   order is from top to bottom: 29 Sep, 4 Oct, and 5 Oct 2000. The beam
   size, total flux density and RMS are given in the observing log,
   Table~\ref{VSOP_tab:obslog}. Intensity profiles through the core component that
   better show the variability are given in Fig.~\ref{VSOP_fig:VSOPprofile}}\label{VSOP_fig:VSOPmaps}
\end{figure*}

Combining the VLBI ground stations with the radio antenna onboard HALCA improves
the resolution at 5\,GHz by a factor of three (see Table~\ref{VSOP_tab:obslog}).
Owing to the small diameter of the HALCA antenna (diameter, $\varnothing=8$\,m),
the sensitivity on the VLBI baselines to the satellite is relatively low,
leading to higher noise on the longest VLBI baselines. The nominal system
equivalent flux density (SEFD) of the HALCA antenna is 15300\,Jy at 5\,GHz
(\citealt{2000PASJ...52..955H}) compared to 312\,Jy for each VLBA antenna
($\varnothing=25$\,m) and 18\,Jy for Effelsberg ($\varnothing=100$\,m). The
uniform weighted $I$ and $P$ images are presented in
Fig.~\ref{VSOP_fig:VSOPmaps}. To achieve the highest possible resolution, equal
weights were applied to all antennas.

In total intensity, the jet extends to core separations of up to 3.5\,mas. In
comparison to the ground array maps, the jet appears less straight, with bends,
followed by back-bends. Although the intensity is weak, the polarization vectors
in the jet seem to follow these oscillations. The VLBI core itself completely
dominates the weak jet. Due to the 3 times higher resolution with HALCA, it is
easier to distinguish between compact emission from the core and extended
emission from the mas-jet. The right panel of Fig.~\ref{VSOP_fig:VSOPmaps} shows
the polarization contour maps. Owing to their relatively low dynamic range of
70:1, only the VLBI core component is detected at a significant level. Most of
the polarized jet emission remains only marginally visible, as indicated by the
grey line, which marks the lowest contour of the emission in total intensity. In
addition to the core region, a region of slightly enhanced jet polarizations is
visible between 1.5\,mas to 2\,mas from the core. As for the ground array maps,
we summarise the integrated flux densities of core and jet in total intensity
and polarization in Table~\ref{VSOP_tab:VSOPresult}. Since the marginal
detection of linear polarization in the jet prevents reliable measurements, we do
not give linear polarization values for the jet in
Table~\ref{VSOP_tab:VSOPresult}.

In comparison to the ground array data (see Table~\ref{VSOP_tab:VLBAresult}),
we measured a core flux density that is $25-35$\,\% lower, but a jet flux
density that was comparable, although the jet is much shorter. This is most
likely due to blending effects between core and jet emission, as we could not
properly separate both regions using the ground array images. In all three VSOP
observations, $\sim 6$\,\% of the total flux density seen in the ground array
maps is missing in the space-VLBI maps. This is reasonable, since the outer
jet is partially resolved and therefore much shorter. It is possible to recover
the missing flux density by using natural weighting, but this would also
decrease the resolution. Since from the ground array images it is already
known that the jet flux density is not variable, we can concentrate in the
following on the analysis of the core-variability with the highest possible
angular resolution.

\begin{table}[htbp]
\centering
\caption{Summary of the flux densities of the VSOP
maps (Note that reliable measurements of the jet's linear polarization were not
possible).}
\label{VSOP_tab:VSOPresult}
\begin{tabular}{lld{4}d{3}d{3}}
\hline
Epoch &  & \multicolumn{1}{c}{I [mJy]} & \multicolumn{1}{c}{P [mJy]} & 
\multicolumn{1}{c}{$\chi$ [$^\circ$]}\\
\hline
\multirow{2}{11pt}{Sep~29} & Core & 492.5 , 25.0    & 11.8 , 2.8 &  49.6 ,\5pt 4.3 \\
                       & Jet  &  62.4 ,\5pt 4.0 &           &              \\
\multirow{2}{11pt}{Oct~4} & Core & 465.4 , 23.6    & 10.8 , 2.6 &  40.3 , 4.1 \\
                       & Jet  &  59.5 ,\5pt 3.7 &           &              \\
\multirow{2}{11pt}{Oct~5} & Core & 462.4 , 23.3    & 5.6 , 1.4 &  54.9 , 4.5 \\
                       & Jet  &  60.1 ,\5pt 3.7 &                          \\
\hline
\end{tabular}
\end{table}

The VSOP images show the same variability behaviour as the ground array images.
The total intensity of the core varied at a marginal level of $\sim 6.5$\,\%.
The flux density of the jet emission remained constant at the 4.8\,\% level.
Again, the linearly polarized flux density of the core was highly variable, with
an amplitude modulation of a factor 2 between October 4 and 5. The integrated
polarization flux densities of the core is comparable to the ground-array
images, which suggests that the linear polarized component itself is still
compact on the VSOP scales ($\leq 0.25$\,mas). We note that 0716+714 was
recently detected with VLBI at 230\,GHz on transatlantic baselines
(\citealt{Krichbaum2004}). This sets an upper limit to the source size of $\sim
20\,\mu$as, an order of magnitude smaller than the resolution obtained by VSOP.

\begin{figure}[htbp]
\centering
\includegraphics[bb=0 0 595 842,angle=0,width=9.5cm]
{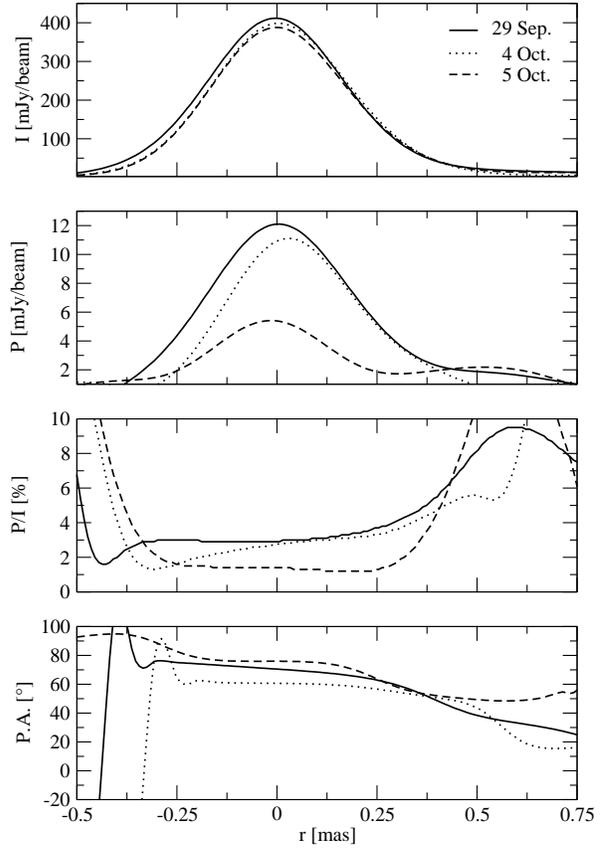}
   \caption{Slices along ${\rm P.A.}=12.5^\circ$ through the VSOP images of
   the core of 0716+714 in total intensity, linear polarization intensity, fractional
   polarization and EVPA. It is a virtual zoom into the core region of
   Fig.~\ref{VSOP_fig:groundprofile}, and $r=0$ marks the same position in both
   plots.}\label{VSOP_fig:VSOPprofile}
\end{figure}

As in the ground array maps, the EVPA in the core of 0716+714 rotates between
the three epochs. Within the measurement errors, the vectors rotated by the same
amount as in the ground array images, namely around $-9^\circ$ between the first
and the second epochs and around $+12^\circ$ between the second and the third
epochs. The source profile of the VSOP-images is shown in
Fig.~\ref{VSOP_fig:VSOPprofile}. For the core region, it looks very similar to
the profiles  from the ground-array images (Fig.~\ref{VSOP_fig:groundprofile}).
The VSOP profiles reveal a small but clearly visible  position  shift of the
linear polarized peak between the epochs. Since these shifts are very small
($<50$\,$\mu$as), we are not confident that they are real. Small position shifts
could be caused by the motion of the components,  which -- with regard to the
short observing time interval (6 days) and the resulting extreme apparent
velocities --  does not appear very likely. An alternative and perhaps more
realistic interpretation could invoke blending effects between two or more
polarized and variable sub-components, located below the angular resolution of
this data sets and inside the core region (see also Sect.~\ref{VSOP_sec:tb}).

\subsection{Results from total flux density measurements}\label{VSOP_sec:Eff}

Table~\ref{VSOP_tab:ebsources} summarises the total flux density measurements,
which were taken between VLBI scans with the 100\,m radio telescope in Effelsberg.
The last two columns of the table give the modulation index $m$ and the reduced
$\chi^2_{\rm r}$ for testing the variability. Comparison of the modulation
indices and $\chi^2_{\rm r}$ of 0836+71, 0615+82, and 0716+714 clearly shows
that 0716+714 was variable during the observations. The formal $\chi^2$-test
yields a probability of significant intra-day variability of 99.984\,\%. The
same arguments apply for the polarization. In Table~\ref{VSOP_tab:ebpol} we
summarise the linear polarization properties of 0716+714 and the polarization
calibrators. The calibrators 0615+82 and 0951+69 are not shown since they are
unpolarized.

\begin{table}[htbp]
\caption{Linear polarization as
observed by Effelsberg. $\langle  \rangle $ indicate mean values.} 
\centering
\scriptsize
\label{VSOP_tab:ebpol}
\begin{tabular}{ld{3}d{2}..d{3}.}
\hline
\multicolumn{1}{c}{Source} &
\multicolumn{1}{c}{$\langle P\rangle $} &
\multicolumn{1}{c}{$\langle P/I\rangle $} &
\multicolumn{1}{r}{$m_{\rm P}$} &
\multicolumn{1}{c}{$\chi^2_{\rm r}$} &
\multicolumn{1}{c}{$\langle \chi\rangle $} &
\multicolumn{1}{c}{$\chi^2_{\rm r}$}\\
&
\multicolumn{1}{c}{[mJy]} &
\multicolumn{1}{c}{[\%]} &
\multicolumn{1}{r}{[\%]} &
 &
\multicolumn{1}{c}{[$^\circ$]} &
\\
\hline
0716+71 &  19.9 ,\hspace{3.5pt}  3.3 &  2.7 , 0.4 &  9.26 & 15.39 &  18.8 ,\hspace{3.5pt}  4.6 & 24.2\\
0836+71 & 160.0 ,\hspace{3.5pt}  1.1 &  6.4 , 0.1 &  0.00 &  0.05 & 106.3 ,\hspace{3.5pt}  0.3 &  0.2\\
3C\,286   & 823.3 ,\hspace{0pt} 12.8 & 11.0 , 0.3 &  0.78 &  0.27 &  33.0 ,\hspace{3.5pt}  0.2 &  0.2\\
\hline
\end{tabular}
\end{table}

\begin{figure}[htbp]
\centering
\includegraphics[bb=34 53 542 781,angle=0,width=9cm,clip] {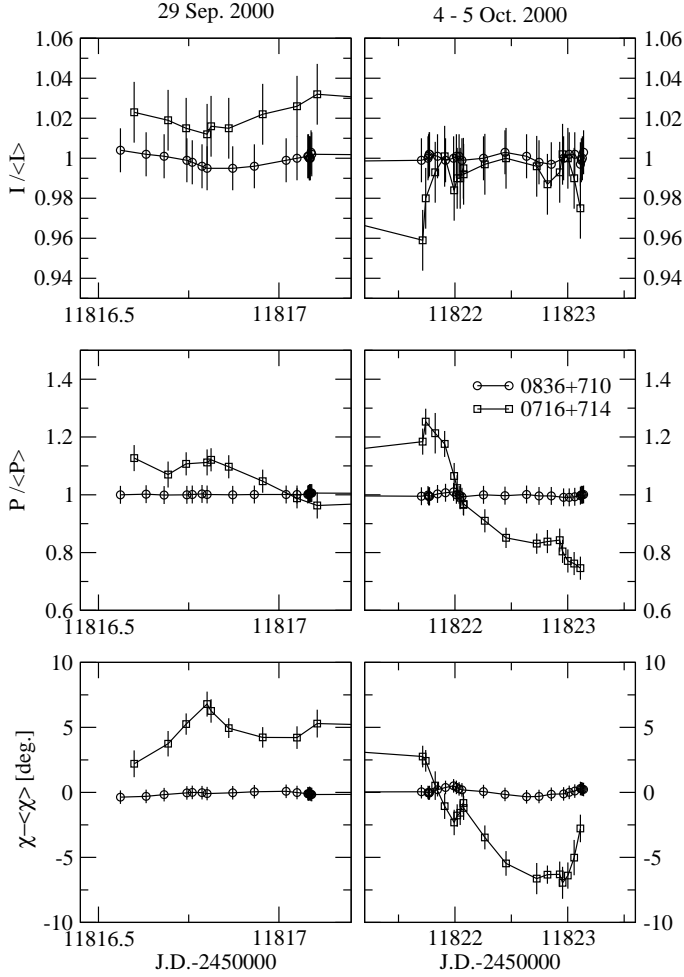}
   \caption{Total intensity (top), linear
   polarization (middle), and EVPA variations of 0716+714 and the calibrator
0836+710, as measured at Effelsberg at 5\,GHz. The left panel shows the data
from September 29 and the right panels from October 4 to 5 (note that the time
axis between left and right panels differ). For intensity and polarization, we
show relative variations, normalised to the mean (average taken over all three
days). For the EVPA we plot the absolute difference relative to the mean EVPA.
The mean values are given in
Table~\ref{VSOP_tab:ebsources}~\&~\ref{VSOP_tab:ebpol}.} \label{VSOP_fig:eblc} 
\end{figure}

Again 0716+714 is the only source that showed significant variability. The total
intensity, linear polarization, and EVPA light curves for 0716+714 and 0836+710
are presented in Fig.~\ref{VSOP_fig:eblc}. Since these measurements were made
primarily for telescope pointing, the accuracy for flux density measurements is
not as high, as it is typically the case for dedicated IDV monitoring
observations. We therefore smoothed the light curves, using a three-point
running mean, in order to reduce the noise. Figure~\ref{VSOP_fig:eblc} clearly
shows the variability of 0716+714 and the non-variability of its nearby
secondary calibrator 0836+710. In 0716+714, up to 5\,\% variations are seen in
total intensity and about 40\,\% in linear polarization. The polarization angle
changed by up to 10 degrees.

\section{Discussion}\label{VSOP_sec:discussion}

Here we compare the intra-day variability seen in total intensity and linear
polarization with the Effelsberg telescope to the variations seen in the VLBI
images and discuss then their implications for the origin of the variability
(Sect.~\ref{VSOP_sec:origIDV}~\&~\ref{VSOP_sec:multcompos}) and for the
brightness temperature of the VLBI core of 0716+714 (Sect.~\ref{VSOP_sec:tb}).
We also discuss possible physical causes for our findings. Using the results of
a recent VLBI study of the jet kinematics of 0716+714 (B05), we were able to
derive some basic jets parameters, like speed, Doppler-factor, and inclination
(Sect.~\ref{VSOP_sec:jetparameter}).

\subsection{Origin of variablity}\label{VSOP_sec:origIDV}

A simple comparison of the average flux density measurements from Effelsberg and
the integrated flux densities of the core and the jet region at the three VLBI
epochs reveals that all the variablity originates in the VLBI core component
(see Table~\ref{VSOP_tab:VLBAvsEb}). The mean difference between the total flux
density in the VLBI maps and the single-dish flux density is
($183.4\pm3.1$)\,mJy, corresponding to about 24.5\,\% less flux density in the
VLBI images. This is plausible, since the arcsecond-scale structure of 0716+714
shows two radio-lobes and a surrounding halo-like structure with a diameter of
10\,arcsec (\citealt{1986AJ.....92....1A}), which is inside the Effelsberg beam
but not seen on VLBI scales. Within the measurement uncertainties, the absolute
values of the core variability are similar to those seen at Effelsberg. We
observed a decrease of 21\,mJy ($\sim 3$\,\%) with VLBI and 27.5\,mJy ($\sim
5$\,\%) at Effelsberg between epochs one and two and a constant flux density
level between epochs two and three. This also confirms our estimate of about
2.3\,\% for the relative flux density error between the VLBI epochs 
(Sect.~\ref{VSOP_sec:groundarray}).

\begin{table}[htbp]
\centering
\caption{The VLBI results in comparison
with the single-dish
measurements from Effelsberg. The total linear polarization values of the VLBI
images were calculated from the vector sum of the core and the jet polarization
vectors.}
\label{VSOP_tab:VLBAvsEb}
\begin{tabular}{lld{4}d{2}d{3}}
\hline
Instrument & Part & \multicolumn{1}{c}{I [mJy]} & \multicolumn{1}{c}{P [mJy]} & \multicolumn{1}{c}{$\chi$ [$^\circ$]}\\
\hline
\multicolumn{5}{c}{A2: 29 Sep 2000}\\
\multirow{3}{11pt}{VLBI} & Core  & 520.3 , 26.9    & 12.1 , 1.3 &  49.4 , 4.1 \\
   & Jet   &  56.0 ,\5pt 4.7 &  7.4 , 0.8 & -10.8 , 5.6 \\
   & Total & 576.3 , 31.6    & 17.0 , 2.6 &  27.3 , 6.0 \\
Eb &       & 763.2 ,\5pt 6.9 & 21.4 , 2.6 &  23.4 , 2.1 \\
\multicolumn{5}{c}{A4: 4 Oct 2000}\\
\multirow{3}{11pt}{VLBI} & Core  & 499.3 , 26.1    & 11.8 , 1.3 &  40.7 , 4.0 \\
   & Jet   &  54.8 ,\5pt 6.3 &  7.3 , 0.8 & -11.2 , 7.8 \\
   & Total & 554.1 , 32.4    & 17.3 , 2.7 &  21.3 , 4.8 \\
Eb &       & 735.7 , 16.2    & 21.6 , 2.6 &  18.6 , 2.2 \\
\multicolumn{5}{c}{A5: 5 Oct 2000}\\
\multirow{3}{11pt}{VLBI} & Core  & 503.9 , 25.4    &  6.5 , 1.1 &  52.7 , 5.2 \\
   & Jet   &  54.7 ,\5pt 6.0 &  7.5 , 0.8 &  -9.5 , 7.4 \\
   & Total & 558.6 , 31.4    & 12.0 , 2.3 &  19.1 , 5.3 \\
Eb &       & 740.2 , 14.6    & 15.7 , 1.1 &  13.3 , 2.5 \\
\hline
\end{tabular}
\end{table}

The variablity of the polarized flux density is anti-correlated with the total
intensity variation. The core polarization shows nearly no variation between the
first two epochs and a decrease of 5.3\,mJy in the VLBI images and 5.9\,mJy at
Effelsberg between epochs two and three, which corresponds to a 40\,\% decrease.

Only $4.2\pm0.3$\,mJy of linear polarization are missing in the VLBI maps in
comparison to the $\sim180$\,mJy missing in total intensity. This yields an
average fractional polarization for the large-scale structure of only 2.3\,\%,
which is small when compared to the $\sim40$\,\% linear polarization found with
the VLA at 1.4\,GHz in the lobe-like structure (\citealt{1987MNRAS.228..203S}).
However, the VLA image shows a variety of EVPA orientations, and the discrepancy
is most likely due to in-beam depolarization at Effelsberg, so that polarization
vectors at opposite position angles cancel. The missing polarization flux
density also causes a difference between the EVPA measured from the VLBI images
and at Effelsberg ranging from $2.7^\circ$ to $5.8^\circ$, but an inspection
of the absolute differences between the epochs reveals that the VLBI-core EVPA
varied in a similar manner to the EVPA of the Effelsberg data.

\begin{figure}[htbp]
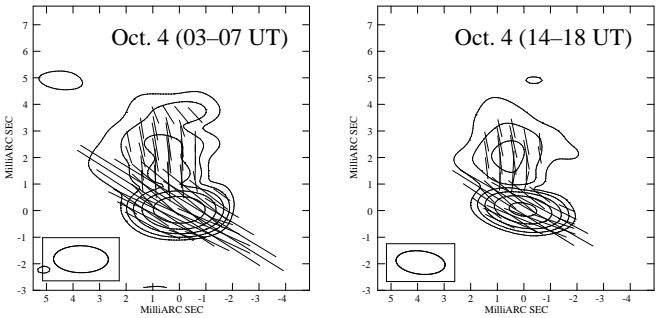

\centering
\hbox{\hspace*{-2mm}
\includegraphics[angle=-90,width=4.3cm]
{3943_fig23.ps}\hspace*{2mm}
\includegraphics[angle=-90,width=4.3cm] {3943_fig24.ps}
 \put(-175,-16){\makebox(0,0){Oct. 4 (03--07 UT)}}
 \put(-45,-16){\makebox(0,0){Oct. 4 (14--18 UT)}}}
   \caption{Linear polarization contour images of the split Oct. 4 VLBI data with
   polarization E-vectors superimposed. At a resolution of about 2\,mas $\times$
   1\,mas at $\sim 85^\circ$, the contours start at 0.5\,mJy/beam, increasing
   in steps of 2.} \label{VSOP_fig:oct4vlbi} 
\end{figure}

Unfortunately, the time sampling of the Effelsberg light curve was not dense
enough to measure reliable intra-day variability for most of the time. However,
on Oct. 4 the light curve showed a significant decrease in polarization from
$(24.5\pm1.7)$\,mJy between $03^{00}$ UT and $07^{00}$ UT to $(20.1\pm1.1)$\,mJy
between $14^{00}$ UT and $18^{00}$ UT. At the same time the polarization vector
rotated by $(4\pm2)^\circ$. To check whether this polarization IDV is also seen
in the VLBI data, the data were split into the corresponding time intervals and
imaged separately. The resulting maps show a decrease in the linear polarization
in the core from ($13.5\pm1.4$)\,mJy to ($9.5\pm1.0$)\,mJy
(Fig.~\ref{VSOP_fig:oct4vlbi} and Table~\ref{VSOP_tab:polIDV}).

\begin{table}[htbp]
\centering
\caption{Linear
polarization variability from two time intervals of VLBI data (October 4th) in
comparison to the single-dish measurements from Effelsberg.} 
\label{VSOP_tab:polIDV}
\begin{tabular}{ld{4}d{4}d{4}d{4}}
\hline
UT [h] &
\multicolumn{1}{c}{$P_{\rm Eb}$ [mJy]} & 
\multicolumn{1}{c}{$P_{\rm VLBI}$ [mJy]} &
\multicolumn{1}{c}{$\chi_{\rm Eb}$ [$^\circ$]} &
\multicolumn{1}{c}{$\chi_{\rm VLBI}$ [$^\circ$]}\\
\hline
03--07 & 24.5 , 1.7 & 13.5 , 1.4 & 21.0 , 2.3 & 40.8 , 4.2 \\
14--18 & 20.1 , 1.1 & 9.5  , 1.0 & 17.2 , 1.8 & 43.0 , 5.1 \\
\hline
\end{tabular}
\end{table}

Although the inaccuracy of the core EVPA measurements in the VLBI images is too
large to also confirm a rotation of the EVPA, the very good agreement between
the intensity variation from hours to days between the VLBI-core and the
single-dish measurements leaves no doubt that the IDV is coming from the VLBI
core of 0716+714. 

We note that similar variations were previously observed in the IDV sources
0917+624 and 0954+658, where ground-based VLBI observations revealed significant
polarization IDV on mas-scales, without showing large variations in the total
intensity (\citealt{2000MNRAS.315..229G}).

\subsection{Multiple components}\label{VSOP_sec:multcompos}

The only moderate variability of the total intensity, the relatively large
variation of the  polarized flux density, and the relatively small variation of
the polarization angle indicate that the variations are caused by the vector sum
of the variability of multiple compact and differently polarized sub-components,
which are located within the VLBI core region on scales that are smaller than
the beam size. The slope of the polarization angle over the core region as seen
with the higher resolution from the VSOP data (see
Fig.~\ref{VSOP_fig:VSOPprofile}) supports the idea that the VLBI core component
itself is a composite of several sub-components. Such a multi-component
structure is required to explain the polarization IDV by refractive interstellar
scintillation (RISS) (e.g.\
\citealt{1995A&A...293..479R,2001Ap&SS.278....5R,2001A&A...367..770Q}), but is
also indispensable in models using solely jet intrinsic (e.g.\
\citealt{1993ChA&A..17..229Q}) or mixtures of intrinsic and extrinsic effects
(\citealt{2002ChJAA...2..325Q}).

If interstellar scintillation is the dominant cause of the variability of the
sub-components, their size cannot be much larger than the scattering size of the
interstellar medium, which at 5\,GHz and towards 0716+714 is believed to be of
the order of several ten micro-arcseconds (e.g.\ \citealt{1995A&A...293..479R}).
Under these circumstances, the number of sub-components located within the
VLBI-core region (smaller than about 100\,$\mu$as, see next section) is limited
to a few. The independent and possibly partly quenched scintillation
(\citealt{1986ApJ...307..564R}) of the  sub-components could then explain the
variability of the polarization intensity and the variation in the polarization
vector. The relative strength of the variability of the polarized flux density
($\sim 40$\,\%) suggests (i) that at least one of the sub-components is
considerably smaller than the scattering size (strong scintillation), and (ii)
that the number of independently variable components is not very large,
otherwise the time averaged vector sum of the polarization would cancel. Since
the polarization angle variations are small ($\leq 15^\circ$), we conclude that
either (i) the polarization vectors of the sub-components are more or less
aligned or (ii) that the more variable sub-component has to be less polarized
than the less variable (and most likely more extended) sub-component.

It is tempting to identify the smaller and less polarized sub-component with the
optically thick jet-base and the larger and more polarized sub-component with
the optically thin inner jet. The fact that the total intensity is less variable
than the polarized intensity indicates that the polarized sub-components are
considerably smaller than the component(s), which dominate in total intensity.
Of course, this simple scenario could change, if a larger number of
independently varying sub-components exist or if some sub-components vary in a
coherent manner. It is also possible that the variability of the sub-components
is not due to scintillation, but is source intrinsic or, more likely, a mixture
of both effects.

\subsection{Brightness temperature}\label{VSOP_sec:tb}

Assuming that the variability has, at least partly, an intrinsic origin
restricts the emitting region to being extremely compact, which then results in
very high brightness temperatures exceeding the inverse-Compton limit of 
$10^{12}$\,K (\citealt{1969ApJ...155L..71K}). The short time-scales observed
imply linear dimensions of

\begin{equation}
l=\frac{c\,\delta\,\Delta t}{1+z},
\end{equation}
where $\delta$ is the Doppler factor. Assuming stationarity and a redshift of
0.3, the size of the emission region responsible for the variability is
around $8\cdot 10^{12}$\,m or $\sim 55$\,A.U., which is slightly larger than our
solar system. The brightness temperature for a stationary source is
given by 

\begin{equation}
T_{\rm b} = 3.06\times 10^8\ S\ \left(\frac{d_{\rm L}}{\nu\ t\ (1+z)^2}\right)^2 
\end{equation} 
where $S$ (in Jy) is the flux density, $d_{\rm L}$ (in Mpc) the luminosity
distance, $\nu$ (in GHz) the observing frequency, $t$ (in yr) the variability
timescale, and $z$ the redshift. The luminosity distance was
calculated by adopting an analytical fit by \cite{1999ApJS..120...49P} resulting in
$d_{\rm L}=1545.8$\,Mpc (see also B05). To calculate the variability time-scale
we use:

\begin{equation}
t=\frac{\langle S\rangle }{\Delta S}\ \frac{\Delta t}{(1+z)}, 
\end{equation} 
where $\langle S\rangle $ (in Jy) is the mean flux density, $\Delta S$ (in Jy) the
standard deviation, and $\Delta t$ (in yr) the duration of the variation (e.g.
\citealt{1996AJ....111.2187W}). Calculating the corresponding $T_{\rm b}$ of the
linear polarization decrease between October 4th and 5th ($\langle S\rangle =9.2$\,mJy,
$\Delta S=3.8$\,mJy, and $\Delta t=24$\,h) results in $T'_{\rm
b}\approx3.8\times10^{15}$\,K and the decrease during October 4th
($\langle S\rangle =11.5$\,mJy, $\Delta S=2.8$\,mJy, and $\Delta t=10$\,h) yields $T'_{\rm
b}\approx9.6\times10^{15}$\,K. Since the observed brightness temperature
$T'_{\rm b}$ derived from variability is connected to the intrinsic brightness
temperature by $T'_{\rm b}=T_{\rm b}\cdot\delta^3$, a Doppler factor of 14 to 22
is needed to reduce the brightness temperature to the inverse-Compton limit.
These calculations usually require a densely sampled  light curve to estimate
$\langle S\rangle $, $\Delta S$, and $t$. However, considering that, due to the
insufficient sampling we, have probably not observed the full amplitude or
missed the most rapid variations, the calculated values of $T_{\rm b}$ are
underestimates rather than overestimates of the true values. 

If instead of the inverse-Compton limit, the equipartition brightness
temperature of $5\times 10^{10}$\,K to $10^{11}$\,K is used (cf.
\citealt{1994ApJ...426...51R}), a Doppler factor of up to 60 is obtained. However,
only a small departure from equipartition, e.g. by a factor of two, would  bring
the Doppler factor back to 30, which is well within the range of possible 
Doppler factors derived from our kinematic study (B05).

Gaussian modelfits to the VSOP $(u,v)$-data of the first epoch reveals a flux
density of ($0.42\pm0.06$)\,Jy and a size of $(0.09\pm0.02$)\,mas for the core
component. Since the core is not resolved, the size only represents an upper
limit. However, this measurement already yields a brightness temperature of
$(2.6\pm0.7)\times 10^{12}$\,K and therewith exceeds the inverse-Compton limit.
To bring these temperatures down, a Doppler factor of $> 4$ is needed. We note
that all these values depend on the redshift of 0716+714 which is not yet known.
The used value of  $z=0.3$ is based on the non-detection of an underlying host
galaxy (\citealt{1996AJ....111.2187W}), but newer studies already place the
limit to $>0.5$ (\citealt{Sbarufatti2005}) which would increase the observed
$T'_{\rm b}$ by at least a factor of three. A lower limit of the Doppler factor
of 2.1 comes from synchrotron self-Compton (SSC) models
(\citealt{1993ApJ...407...65G}).

The VSOP 5\,GHz AGN survey (\citealt{2000PASJ...52..997H}), which contains
nearly all extra-galactic flat-spectrum radio sources brighter than 1\,Jy and at
galactic latitude  $\geq 10\,^\circ$ ($\sim 300$ sources), has shown that about
50\,\% of these types of sources have brightness temperatures of $T_{\rm
b}>10^{12}$\,K, and about 20\,\% even have $T_{\rm b}>10^{13}$\,K
(\citealt{2004ApJS..155...33S,2004ApJ...616..110H}). The VSOP observations of a
sub-sample ($\sim 30$ sources) of the Pearson-Readhead sample
(\citealt{1988ApJ...328..114P}) revealed a significant correlation between the
IDV activity and the brightness temperature (\citealt{2001ApJ...549L..55T}). The
authors find that sources showing rapid IDV have higher brightness temperatures
than weakly variable or non-IDV sources. Therefore, independent of the physical
model for the variability, relativistic beaming seems to play a role.

\subsection{Jet inclination and Lorentz factor}\label{VSOP_sec:jetparameter}

Simulations of relativistic hydrodynamic jets show that shocks are likely to
occur in highly supersonic flows. Those provide a natural way to locally enhance
the magnetic field and relativistic electron density, producing knots of
emission such as seen in VLBI images (e.g., \citealt{1997ApJ...482L..33G}, and
references therein). Assuming that the knots in the jet are shocks travelling
down the jet, we can use the observed degree of polarization of such a jet
component to set limits on the angle to the line of sight and on the jet Lorentz
factor (\citealt{1988ApJ...332..696C}). The shock model used here is based on
the scenario that \cite{1985ApJ...298..301H} used to describe the polarization
variability in BL\,Lac. In that model, a propagating shock compresses the
magnetic field of the jet so that the initially random field is ordered in a
plane transverse to the jet axis. With this we can constrain  the amount of
compression, defined as a unit length compressed to a length $k$, and the angle
between the line of sight and the shock plane in the frame of the jet
($\epsilon$) using the measured degree of polarization ($m$) in the jet. If we
assume that the shock travels with the same velocity as the emitting material,
we can also relate $k$ and $\epsilon$ to the apparent speed of the jet and
constrain the inclination of the jet, $\theta$, and the jet Lorentz factor,
$\gamma$ (\citealt{1988ApJ...332..696C}).

A modelfit to the linear polarization image with JMFIT in {\sc Aips} reveals
that the peak of the linear polarization in the jet corresponds to component C7 
(B05), which moves with a speed of $(6.9\pm0.2)\,c$ assuming $z=0.3$.
Integrating the flux density in the total intensity and in the polarization
image over a region of 1.5 times the beam width around the component centre ($r=
2.7\pm0.2$\,mas) yields a fractional polarization, $m$, of about $(24\pm4)$\,\%.
For a spectrum of $\alpha_{\rm 5/15\,GHz}=-0.7$ (B05) and therewith
$s\approx2.5$, the nominal degree of  polarization of synchrotron emission in a
uniform magnetic field is $m_0\approx0.72$.

We would like to distinguish two cases. First, if  the jet is oriented at the
viewing angle that maximizes the apparent superluminal motion, then $\epsilon=0$
and $k$ gives an upper limit on the amount of compression. We find $k_{\rm min}
= 0.71\pm0.04$. Second, if we assume a maximum of compression ($k\ll 1$), then
$\epsilon$ corresponds to the maximum angle between the shock plane and the line
of sight, which yields $\epsilon_{\rm max} = \pm 45\pm4^\circ$.

In the first case, the measured apparent speed of  $(6.9\pm0.2)\,c$ of component
C7 yields a jet Lorentz factor of $\sim 7$ and an angle to the line of sight of
$\sim 8^\circ$, which would not agree with the results (B05), where a minimum
Lorentz factor of $11.6$ and $\theta_{\rm max}=4.9\,^\circ$ was found. The
second case ($k\ll 1$ and $\epsilon =\pm 45^\circ$) yields $\gamma\approx 10$
and $\theta\approx 2.5^\circ$ or $14^\circ$ (depending on the sign of 
$\epsilon$), but the larger viewing  angle can almost certainly be excluded from
the kinematics (B05). From this the second case seems more likely, and the
derived jet parameters show the same trend as the  parameters ($\gamma>16$ and
$\theta<2^\circ$) derived by B05. The shock scenario also favours  the
small-angle solution, where we have higher jet Lorentz factors and smaller
viewing angles than those derived by maximization of $\beta_{\rm
app}=\frac{\beta \sin\theta}{1-\beta \cos\theta}$. The corresponding Doppler
factor is about 17 for $\gamma\approx 10$ and $\theta\approx  2.5^\circ$.

\section{Conclusions}\label{VSOP_sec:conclusions}

In order to search for the origin of the rapid IDV in 0716+714, a multi-epoch
VSOP experiment was performed in September and October 2000. The ground array
and the VSOP images show a bright core and a jet oriented to the north. The
linear polarization images indicate that the jet magnetic field is perpendicular
to the jet axis. Compared to the jet axis, the electric vector position angle in
the core is misaligned by around 60\,$^\circ$. This is explained either by
opacity effects in the core region or by a curved jet. Jet curvature is
supported by recent high resolution 3\,mm VLBI observations that show the inner
jet structure ($r<0.1$\,mas) at a similar position angle as the EVPA in the core
at 6\,cm wavelength. A misalignment between the parsec-scale and the
kiloparsec-scale structure is typical of BL\,Lacs
(\citealt{1988ApJ...328..114P,1992ApJ...391..589W}) and is also seen in 0716+714
(VLA images can be found in \citealt{1986AJ.....92....1A,2000MNRAS.313..627G};
Bach et al., in prep.). The new VLBI observations suggest that the jet bending
continues down to sub-parsec-scales. Possible explanations for such a
misalignment are a helical jet, which is oriented towards us
(\citealt{1993ApJ...411...89C}), possibly due to Kelvin-Helmholtz instabilities
(e.g., \citealt{1986CaJPh..64..484H,1981ApJ...250L...9H}) or to being driven by
precession (B05; \citealt{2005AJ....130.1466N}), or due to interaction with the
surrounding medium. 

Simultaneous flux-density measurements with the 100\,m Effelsberg telescope
during the VSOP observations revealed variability in total intensity ($\sim
5$\,\%) and in linear polarization (up to $\sim 40$\,\%) accompanied by a
rotation of the polarization angle by up to  $15\,^\circ$. The analysis of the 
VLBI data shows that in 0716+714 the intra-day variability is associated to
the VLBI-core region and not to the milli-arcsecond jet. Both the ground
array and the VSOP maps show a similar decrease of the flux densities in total
intensity and linear polarization of the core component, which is in good
agreement with the variations in the total flux density and polarization seen
with the Effelsberg 100\,m radio-telescope. In this, 0716+714 displays a
behaviour that is similar to what was previously observed in the IDV sources 0917+624 and
0954+658, where components in or near the VLBI core region were also made
responsible for the IDV (\citealt{2000MNRAS.315..229G}). Over the time interval
of our VSOP observations, no rapid variability in the jet was observed and we
cannot confirm the variability outside the core and in the jet found by
\cite{2000MNRAS.313..627G}.

The simultaneous variation of the polarization angle with the polarized
intensity in the core suggests that the variations might be the result of the
sum of the polarization of more than one compact sub-component on scales smaller
than the beam size. Assuming that these variations are intrinsic to the source,
we derived brightness temperatures of $\sim 3\times 10^{15}$\,K to $\sim
10^{16}$\,K. Doppler factors of $>20$ are needed to bring these values down to
the inverse-Compton limit. These numbers agree with the observed speeds in the
jet if the angle to the line of sight is very small ($\theta<2^\circ$), as
already proposed by B05. Because of the unknown redshift, the derived speeds and
brightness temperatures represent only lower limits.

However, interstellar scintillation effects could also explain the IDV seen in
the  VLBI core, if the core region consists of several compact and polarized
sub-components, with sizes of a few ten micro-arcseconds. To explain the
observed polarization variations, the sub-components must scintillate
independently in a different manner, which means that they  must have slightly
different intrinsic sizes and intrinsic polarizations  (i.e.
\citealt{2001Ap&SS.278....5R,2001Ap&SS.278..129R}).

Independent of whether the interpretation of the IDV seen in the VLBI core is
source intrinsic or extrinsic, the space-VLBI limit to the core size gives a
robust lower limit to the brightness temperature of $\geq  2\times 10^{12}$\,K
and therewith exceeds the inverse-Compton limit. This implies a lower limit to
the Doppler factor of about $\geq 4$ and, independent of the model we use to
explain the variability, relativistic beaming seems to play a role.
Unfortunately, the possible sub-components inside the core region cannot be
observed directly. Future multi-frequency polarization VLBI observations,
including simultaneous observations at higher frequencies, should help
distinguishing which fraction of the IDV is due to source intrinsic variations
and which is caused by the interstellar medium.

To explain the enhanced degree of polarization in the superluminal jet,
we applied a simple shock model developed by \cite{1988ApJ...332..696C} to our
observations and found reasonable values for the jet Lorentz factor and the
viewing angle. This supports the idea that the knots of bright emission in the
jet are shocks travelling down the jet (\citealt{1985ApJ...298..114M}).

\begin{acknowledgements}
We thank the anonymous referee
for helpful comments and suggestions. This work made use of the
VLBA, which is an instrument of the National Radio Astronomy Observatory, a
facility of the National Science Foundation, operated under cooperative
agreement by Associated Universities, Inc. This work is
also based on observations with the 100\,m radio telescope of the MPIfR
(Max-Planck-Institut f\"ur Radioastronomie) at Effelsberg. We gratefully
acknowledge the VSOP Project, which is led by the Japanese Institute of Space
and Astronautical Science in cooperation with many organisations and radio
telescopes around the world. U.B. was partly supported by the European
Community's Human Potential Programme under contract HPRCN-CT-2002-00321
(ENIGMA Network).
\end{acknowledgements}


\end{document}